\documentclass[journal]{IEEEtran}
\usepackage{epsfig}
\usepackage{graphics}
\usepackage{amsmath}
\usepackage{amssymb}
\usepackage{multirow}
\usepackage{cite}
\usepackage{array}
\usepackage{pslatex} 
\usepackage{url}
\usepackage{caption}
\usepackage{lineno}
\usepackage{graphicx}  
\usepackage{setspace}
\usepackage{tikz}
\usepackage{hhline}
\usepackage{mathtools}
\usepackage{verbatim}
\usepackage{textcomp}
\usepackage{subcaption}
\ifCLASSINFOpdf
\else
\fi

\begin{document}
%
\title{Characterizing Human Limb Movements Using An In-House  Multi-Channel Non-Invasive Surface-EMG System}
%
%
%

\author{Vinay C K$^1$,
        Madhav Rao$^1$,~\IEEEmembership{Senior~Member,~IEEE},
        and~Vikas Vazhayil$^2$,
        \thanks{
 Vinay C K was an Integrated MTech student at IIIT-Bangalore, and Madhav Rao is a faculty working with IIIT-Bangalore, India. (email:mr@iiitb.ac.in), Vikas Vazhayil is with the Department of Neurosurgery, NIMHANS, Bangalore, India.}}
\maketitle

\begin{abstract}
Electromyography (EMG) signals are obtained from muscle cell activity. 
The recording and analysis of EMG signals has several applications. The EMG is of diagnostic importance for treating patients suffering from neurological
and neuromuscular disorders.
Conventional methods involve placement of invasive electrodes within the muscles to record EMG signals. 
The goal is to showcase the usage of surface based EMG signals to characterize all possible human limb movements.
An in-house non-invasive EMG signal acquisition system that offers  characterization of human limb actions is a suitable candidate for motor impairment studies and easily extendable to design bionics control specifically for neuromuscular disorder patients. 
An in-house 
8-channel surface-EMG signal acquisition system was designed, fabricated, and employed for characterizing specific movements of upper and lower limb. 
The non-invasive acquisition system captures the compound electromuscular activity generated from the group of muscles. 
The EMG acquisition system was designed as a modular structure where
the front end analog circuit designs were replicated for all 8 channels, and were
designed to function independently. 
 Support vector machine (SVM) as classifier models were developed offline to successfully characterize different human limb actions.
The in house built 8 channel acquisition system with ML classifier models were utilized to successfully characterize movements at various joints of the upper and lower limb including fingers, wrist, elbow, shoulder, knee, and ankle individually. The thorough investigation involved optimal number of electrodes for various limb movements.
The movements in the form of abduction, adduction, flexion, extension, supination, and pronation as relevant at all the above mentioned joints were targeted for successful classification.
The proposed EMG system designed in a wearable form demonstrated adequate accuracy and a robust design, envisaging use in diverse setting of normal and abnormal physiology. 
\end{abstract}

%
\IEEEpeerreviewmaketitle

\section{Introduction}
Muscles along with nerve cells shows a unique property of membrane action potential that changes from a resting state to an active state and back. Muscle cell activity in the form of contraction and relaxation
generates electrical signals.
Thus evaluation of electrical activity of muscles provides 
adequate idea of muscle functions. 
Electromyogram (EMG) of skeletal muscles is used for a variety of diagnostic purposes. Important application includes diagnosis and classification of muscle disorders, neuropathies, neuromuscular junction disorders, for neuromonitoring during brain and spinal cord surgery. 
An important area of work is in bionics and robotic devices which mimic limb functions. These devices require input from the normal or diseased limb for controlling the actuation of the device. The robotic devices are meant to improve existing motor power by repetitive exercises and movements~\cite{home, upperlimb1, upperlimb2}. 
Regaining muscle strength through recommended set of exercises and co-ordinated movement is a standard practise in rehabilitative medicine.~\cite{rehab1, rehab2, rehab3}.
Post stroke patients with hemiplegia are generally advised for upper limb and lower limb rehabilitation to aid recovery~\cite{management, management1, management2}. The underlying principle involves regular activation of muscle cells to prevent loss of bulk and atrophy~\cite{running, running2, exercise, exercise2}. 


EMG signals are objectively measured at the surface of the skin and are
considered suitable in the design of rehabilitative devices~\cite{quantitative}.
The EMG signals in the frequency band of 10 to 500 Hz~\cite{frequency-range, frequency-range2}, are generated from the muscle cells, the compound signals across several groups of muscles form the basis of the acquired signal. Generally needle based invasive techniques are the standard method to acquire 
EMG signals, and only recently there is continued interest in surface based EMG (sEMG) signals~\cite{surface}.
The sEMG signals captured non-invasively at the surface, are attenuated, with signal levels falling in the range of tens to hundreds of $\mu$V, however the signal showcase unique characteristics for different muscle activations owing to limb actions.
Hence not only special design elements are required to extract high signal to noise ratio (SNR) sEMG signals, but also a special software algorithm is required to characterize limb movements from the group of extracted sEMG signals.
The sEMG signals amplitude and frequency parameters although marred with artifacts and external noise interference, yet it contains sufficient real time information representing the motor functionality due to muscle activity~\cite{realtime, artifact}.
Hence comprehensive information towards motor recovery
is offered by the sEMG signal for post stroke patients undergoing 
rehabilitation~\cite{motor-neuron, poststroke}.
The premise of sEMG also offers the possibility of intermediate clinical grade assessments to monitor progression and recovery of motor impairment. This in turn is likely to improve the planning of physiotherapy routines. A possible novel application would be home based evaluation using a sEMG device which can help in improved diagnosis and follow up. 

Several wearable devices to measure normalized sEMG signal, are commercially available~\cite{myoband, natus, roam, neurosoft, delsys, cometa}, and are used for various applications including unsupervised keyboard learning~\cite{sharmilaICPR}, and other gesture driven human computer interactive application~\cite{myoband1}. 
The variety of devices are either bulky in nature, rigid in usage, short in supply, difficult to procure, and expensive.
Hence an in-house generic and personal acquisition system with multi-channel measurement capability, allowing spatial freedom to position electrodes and characterize all parts of upper and lower limb actions including movements of fingers, wrists, elbow joint, shoulder joint, knee joint, and ankle joint individually, was decided. The goal of this paper is to demonstrate the usage of surface based EMG signals to characterize all possible human limb movements, which is not comprehensively explored anyweher else, as per the authors' knowledge.

The paper utilizes the system design presented in~\cite{vinay-sEMG} to develop classifier techniques derived from a machine learning (ML) algorithm for characterizing different parts of upper and lower limb actions for the  first time. The application of machine learning techniques(ML) and back propagation neural network (BPNN) algorithms for classifying five upper limb actions including extension, flexion of wrist, open, close, and relax state
of hand with high accuracy using five channel system has been reported earlier~\cite{emgpattern}. The designed EMG acquisition system was confined to forearm to detect wrists and hand postures only, but the system does not showcase
the ability to track independent finger movements which is vital for designing feedback enabled bionics and  rehabilitative robot design. 
Besides, the designed system was not feasible to comprehensively investigate other parts of the upper and lower limb of the body. 
Different classifier techniques in the form of Linear Discriminant Analysis (LDA), Random forest (RF), Principle Component Analysis (PCA), and Support Vector Machine (SVM) were applied for upper limb actions reporting acceptable accuracy~\cite{sapsanis2013,sapsanis201306, sapsanis201307, saponas2008, Saponas2009}.
However the spatial positioning, and optimum number of electrodes required to track actions were not studied entirely. The challenge was to acquire low voltage sEMG signals from a generic, wearable and robust system, and characterize different joint movements of the human upper and lower limbs independently.
Characterizing independent and individual movements of fingers, wrist, elbow, shoulder, ankle, and knee joints are considered effective
for measuring progress during rehabilitation therapy. Hence the aim of this work was to design a general sEMG acquisition module to cover maximum movements of the upper and lower limb joints.
Several ML algorithms were adopted in the past~\cite{feature}, which utilizes conventional signal processing techniques to extract feature vectors, on which the ML algorithms were later applied.
The paper presents 8 channel sEMG acquisition system with an ML classifier model developed from sEMG signal measured at different parts of upper limb, and lower limb and was further validated to 
characterize particular actions of the limb.
The paper comprehensively investigates and distinguishes linear, lateral, and angular actions of upper and lower limb, using an in house sEMG acquisition system, reporting high accuracy results. 

\section{DESIGN}
\begin{figure}
    \includegraphics[width=\linewidth]{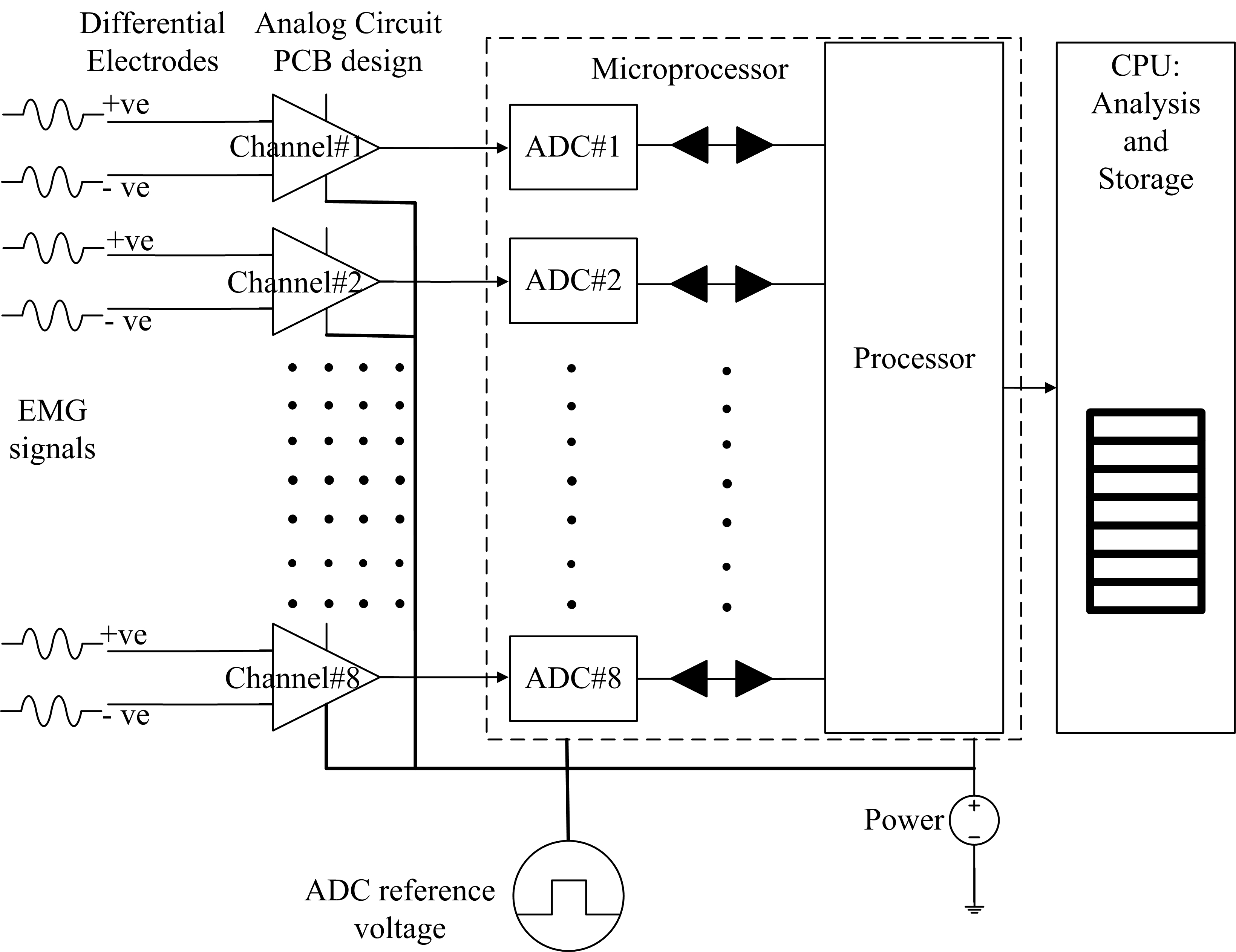}
    \caption{Block diagram representation of 8-channel sEMG signal acquisition system.}
    \label{fig:block}
\end{figure}

A signal acquisition circuit was desired to capture raw sEMG signals from
multiple locations, 
and spatially optimize the electrode positions for accurately characterizing the limb movements. Hence multiple and isolated channels that have the independent capability to acquire sEMG signals simultaneously, was designed to develop a classifier model and further utilize the same for characterizing different actions.
The modular acquisition system for 8 channels were designed on PCB, and individual channels were wired to the electrodes pairs, such that on use case basis, one can deploy and acquire signals from only relevant number of channels. 
The overall signal acquisition system is shown in the Figure~\ref{fig:block}, where the differential pair of electrodes were
utilized to feed the raw acquired EMG signal to the printed circuit board (PCB) designed analog circuitry. The analog front end PCB design consists of amplifier, filters, and buffer units that maximizes the
Signal to noise ratio (SNR) and avoids any loading effects from the microprocessor inputs. 
The microprocessor with an inbuilt 8 analog to digital converter's (ADC) were employed to digitize sEMG signals for further analysis and development of a classifier model to characterize the limb actions.

\begin{figure}
    \includegraphics[width=\linewidth]{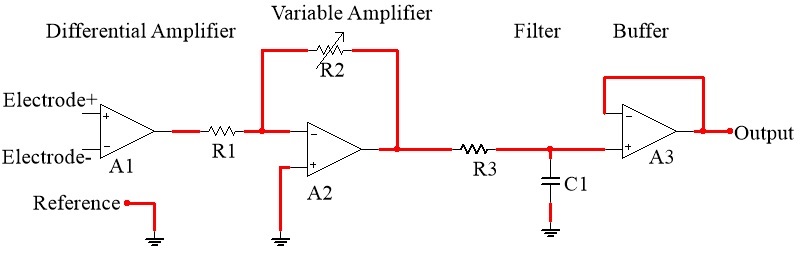}
    \caption{Circuit diagram showing single channel sEMG signal conditioning design.}
    \label{fig:circuit}
\end{figure}

Individual signals captured from the pair of electrodes are conditioned  
appropriately for further digitization.
The sEMG signals are 
amplified using differential amplifier configuration circuit which is later filtered for a specific range of frequency as shown in the
Figure~\ref{fig:circuit}. 
For the signal conditioning circuit design, bipolar configuration was preferred over unipolar architecture, considering improved common mode noise rejection capability, and the spatial freedom to 
acquire sEMG signal along the length of individual muscle fibres, which otherwise is not possible in unipolar configuration.
Gel based adhesive electrodes of type 3M Red Dot Sticky Gel-2560 costing around 10 cents were employed in the designed acquisition system.
The gel electrodes were directly patched to the skin of target actions, with no  preparation. The individual adhesive gel electrodes were electrically connected to the front end PCB design via alligator clips through wires. The differential wires were twisted and designed to shorter length, to shield 
sEMG signal from picking any noise from electrical setup.
A buffer circuit was included to isolate any loading effects on the filter circuits from the microprocessor inputs.
The PCB design ensured that all 8-channel signal conditioning circuit
were housed in a miniature size compactly and also offered
portable and plug-and-play form factor for the signal acquisition system. The PCB signal conditioning circuit was connected to microprocessor to store sEMG signals on one side, and on other 
side, the inputs were plugged to gel electrodes.
The first stage of signal conditioning PCB design referred to $A1$ in the Figure~\ref{fig:circuit},
consists of instrumentation amplifier 
of INA106U/2K5 type from INA106 series to meet the desired 
specification of 
amplified differential signal with fixed gain, but at the same time, apply high rejection ratio for eliminating 
common mode noise signal. 
The second stage consists of a variable gain amplifier 
referred to $A2$ in the Figure~\ref{fig:circuit}, which was
designed from
LM358DT and controlled by a potentiometer. The variable gain amplifier allows sEMG signal level adjustment to overcome highly sensitive nature of the signal
over user anatomy, muscle strength of individual, and skin dryness due to ambient conditions.
The output of the variable
gain amplifier was passed to active high pass filter 
referred to $A3$ in the Figure~\ref{fig:circuit}, configured 
for 30 Hz $f_{3db}$ frequency which involved a buffer designed from LM358DT to 
isolate any loading effects from the in-built ADC of the microprocessor, and filter out any low frequency mechanical perturbations and ambient noise signals.
The buffered analog signal from all 8 channels was further supplied to 
individual ADC's for digitizing the acquired signals.
The 8 channel acquired sEMG signals were sampled at 20 KHz frequency to record the acquired signal in digital form and perform offline classifier model development for characterizing limb movements.



\begin{figure}
        \includegraphics[width=\linewidth]{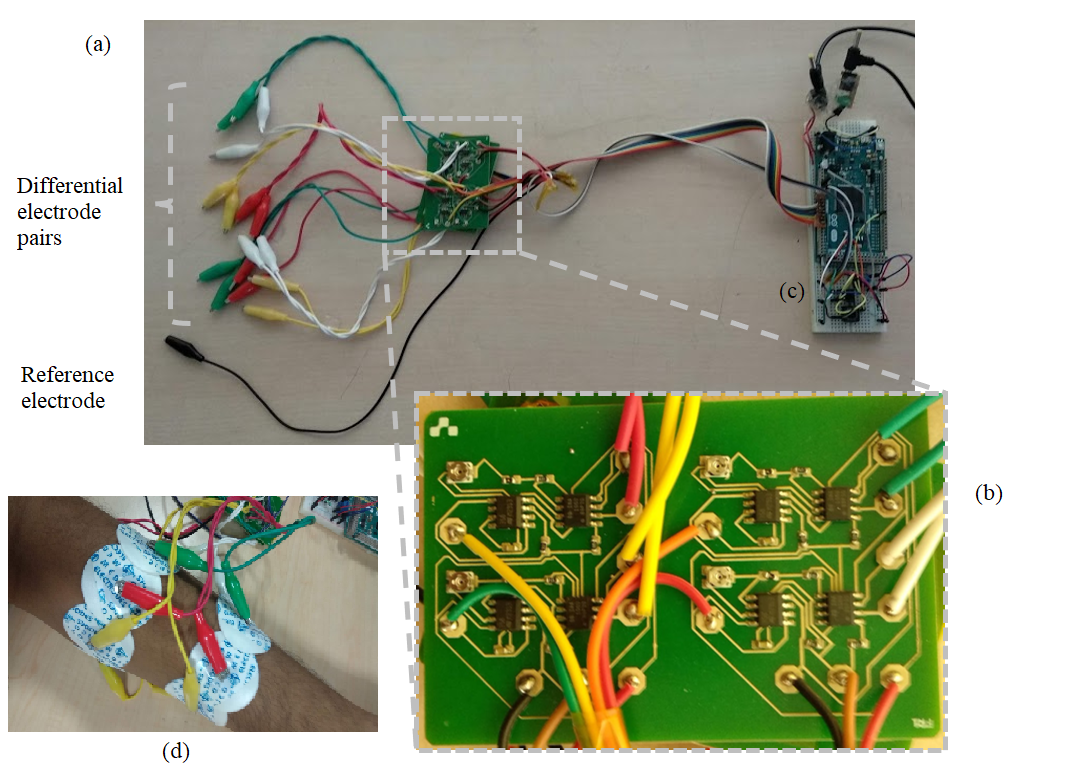}
        \caption{Prototype showing (a) sEMG signal acquisition system with electrodes, (b)~Image of sEMG printed circuit board,
        (c) Image of microprocessor and ADC setup that is interfaced to the acquisition system, and (d) Snapshot of three pairs of electrodes covering posterior side of the forearm.}
        \label{fig:setup}
\end{figure}

A picture of the 8 channel sEMG signal acquisition system including signal conditioning and digitizing is shown in Figure~\ref{fig:setup}(a).
Additive PCB printer was utilized to fabricate the PCB with conductive ink tracers and then appropriate SMD components were reflowed to complete the PCB circuit as shown in Figure~\ref{fig:setup}(b).
Atmel SAM3X8E ARM cortex-M3 microprocessor was integrated in the acquisition system. A reference ADC signal was adjusted in the microprocessor to convert all the 8 analog signals to digital, and is further sampled at a predefined 
20 KHz sampling rate. 
A snapshot showing ARM cortex microprocessor and potentiometer to adjust 
ADC level is shown in the Figure~\ref{fig:setup}~(c).
The microprocessor applies two digital filters: first one notch filter to remove 50 HZ AC power driven noise signals, and second one, a band pass filter with cutoff frequencies designed at 30 Hz and 300 Hz on lower and
higher side of the spectrum respectively, to remove any other 
interfering signals outside the specified band.
The digitally processed time varying oscillating EMG signals with
varying amplitude levels was passed through a continuously moving window Root mean squre (RMS) calculator 
to estimate energy of the signal at discrete times, which is a standard processing practice~\cite{rms-window}.
RMS processed signal quantifies the muscle fibre contraction in the body, which is a primary factor for limb movements, hence windowed RMS level as a feature was utilized to build the classifier model~\cite{rms1, rms2, rms-window}.
The RMS component of the signal from various channels acquired from 
different spatial positions around muscles groups furnish
sufficient indicators 
to characterize limb movements.
This paper does not explore different features sets and different evolving machine learning algorithms, but  addresses the usage of surface EMG signals with primitive ML model to capture all human limb movements. The work 
employs RMS component as the single feature vector and showcases adequate accuracy to characterize different
human limb movements including all joint actions of upper limb and lower limb,
on acquiring surface EMG signals.

\section{Experiments}
\begin{figure}
    \begin{center}
    \begin{subfigure}[b]{0.23\textwidth}
    {\begin{center}
    \includegraphics[scale = 0.17]{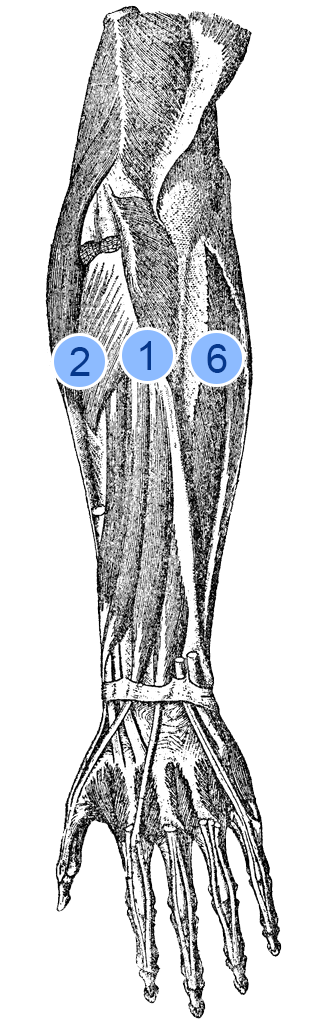}
    \end{center}}
    \caption{}
    \end{subfigure}
    \begin{subfigure}[b]{0.23\textwidth}
    {\begin{center}
    \includegraphics[scale = 0.17]{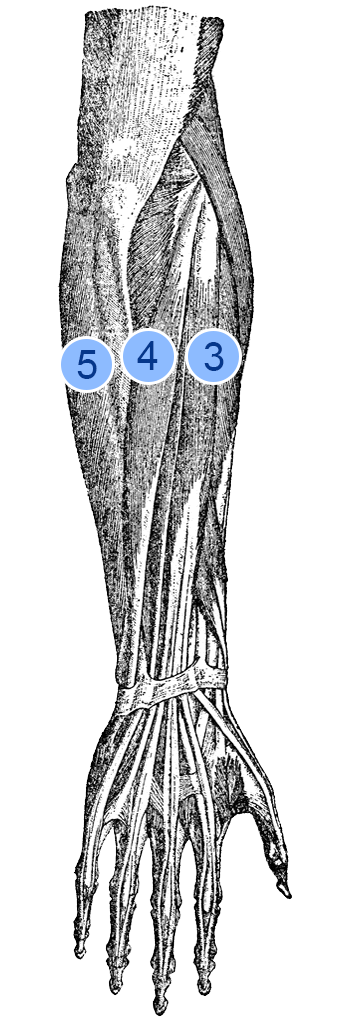}
    \end{center}}
    
    \caption{}
    \end{subfigure}
    
    \begin{subfigure}[b]{0.23\textwidth}
    {\begin{center}
    \includegraphics[scale = 0.23]{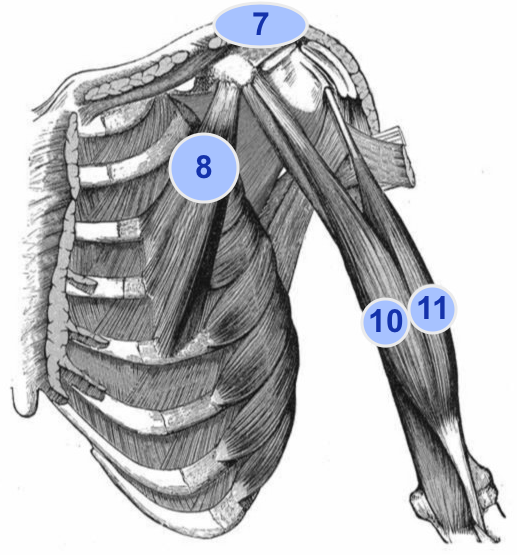}
    \end{center}}
    
    \caption{}
    \end{subfigure}
    \begin{subfigure}[b]{0.23\textwidth}
    {\begin{center}
    \includegraphics[scale = 0.22]{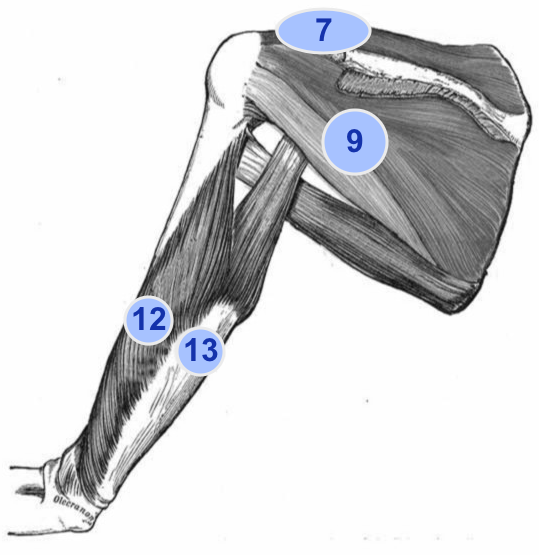}
    \end{center}}
    
    \caption{}
    \end{subfigure}
    \caption{Schematic showing the placement of electrode pairs on upper limb with (a) three electrode pairs on the posterior side of the forearm, (b) three electrode pairs on the anterior side of the forearm. (c) four electrode pairs on the anterior side of shoulder and upper arm, and (d) three electrode pairs on the posterior side of shoulder and upper arm. Upper limb skeletal structure was redrawn from~\cite{armfig}. }
     \label{fig:arms}
    \end{center}
\end{figure}

The designed and fabricated sEMG setup was applied to record 
the following:
wrist and finger actions, 
movements for elbow and shoulder joints, and
movements of knee and ankle joints.
Characterization of individual finger flexion actions and wrist flexion actions, at different forearm postures was intended from the acquired sEMG signals.
The upper limb study included linear and angular actions of elbow joint
followed by
linear, lateral, and rotational actions of the shoulder joint.
Characterization of lower limb was limited to ankle and knee joints, which were considered effective for first level rehabilitation exercises.
A separate characterization and discussion on comprehensive actions offered by the hip joint was required and hence was not included in this study.
The stated upper and lower limb actions were recorded multiple times on a 10 healthy subjects 
between the age group of 18 to 23 years old to validate the working of prototype. 
Consent was taken from all the subjects with a declaration approving that the subjects had no prior neuromuscular weakness or injury, nor any prior surgery performed in the past. 
The research was carried out following the principles of the Declaration of Helsinki.
In addition to this, all methods to acquire signals 
from the subject were performed in accordance with the relevant guidelines and regulations.
All the acquired sEMG data was made freely available in~\cite{github}.
The sEMG acquisition system was applied on the subject and was repeated many times on multiple days to arrive at a robust classifier model.
The subject was asked to relax for 15 minutes before the start of every experiment, to minimize any muscle fatigue. The sEMG acquisition system was applied on the subject right hand and right leg sequentially.
Each experiment of limb movement lasted for 15 seconds with initial relaxation of 5 seconds. 
Six from the eight designed channels were mounted radially on the forearm 
at positions 
referred to as \#1 to \#6, as shown in the Figure~\ref{fig:arms}~(a, b), for classifying fingers and wrist movements. 
A snapshot of three electrode pairs covering the posterior side
of the forearm is shown in the Figure~\ref{fig:setup}~(d). The other 3 pairs of electrodes were symmetrically placed on the anterior side of the
forearm.
A pair of electrodes constituting a channel were placed along the longitudinal dimension of the forearm around the labelled positions with a gap of less than 8~mm in between.
Six pairs of electrodes with three pairs each on anterior and posterior side
completely 
covered the radial circumference of the forearm to measure the electro-muscular signal.
The individual fingers of upper limb were actuated to perform flexion action, and the signal response was acquired by the designed sEMG acquisition system. Similar finger flexion action was repeated at three different rotated positions of the wrists: 0\textdegree, 90\textdegree, and 180\textdegree, where 0\textdegree represented supination pose of the forearm, 180\textdegree represented
pronation pose, and 90\textdegree formed the in-between pose.
Each experiment was repeated 60 times 
to collect adequate and consistent data for developing a 
classifier model.

In a separate experiment, seven differential electrode 
channels were placed 
across the upper arm and the shoulder, which are referred to \#7 to \#13 as shown in the Figure~\ref{fig:arms}~(c,d). 
The elbow joint was moved to perform forearm extension and flexion actions, by keeping upper arm fixed,
and signals were acquired for 60 repeated experiments. Additionally, forearm supination and pronation actions were performed and signals were acquired repeatedly.
In another experiment for shoulder joint, flexion, extension, abduction, adduction, pronation, and supination actions were performed and extracted
sEMG signals were recorded appropriately for further classification.

    \begin{figure}
    \begin{center}
    \begin{subfigure}[b]{0.23\textwidth}
    {\begin{center}
    \includegraphics[scale = 0.38]{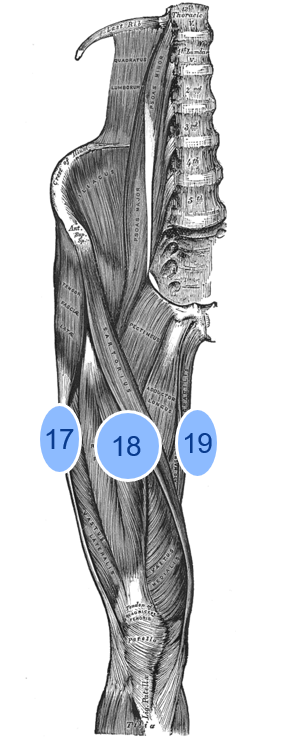}
    \end{center}}
    
    \caption{}
    \end{subfigure}
    \begin{subfigure}[b]{0.23\textwidth}
    {\begin{center}
    \includegraphics[scale = 0.38]{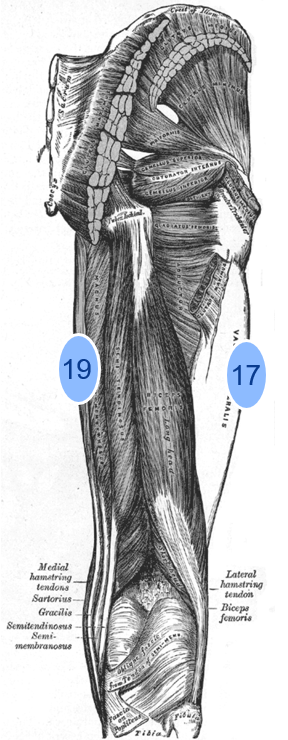}
    \end{center}}
    
    \caption{}
    \end{subfigure}
    \begin{subfigure}[b]{0.20\textwidth}
    {\begin{center}
    \includegraphics[scale = 0.38]{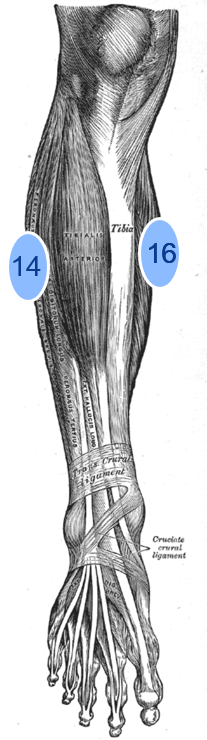}
    \end{center}}
    
    \caption{}
    \end{subfigure}
    \begin{subfigure}[b]{0.20\textwidth}
    {\begin{center}
    \includegraphics[scale = 0.38]{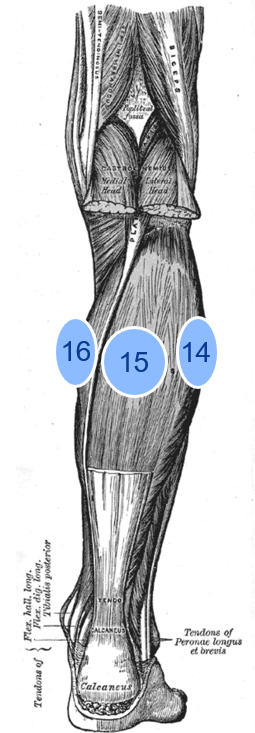}
    \end{center}}
    
    \caption{}
    \end{subfigure}
    
    \caption{Schematic showing placement of electrodes pairs with (a) three electrode pairs on the anterior side of the upper leg, (b) two electrode pairs on the posterior side of the upper leg. (c) two electrode pairs on the anterior side of lower leg, and (d) three electrode pairs on the posterior side of lower leg. Lower limb skeletal structure was redrawn from~\cite{armfig}. }
     \label{fig:feet}
    \end{center}
\end{figure}

An experiment on ankle movement, 
and knee joints 
were performed by placing 
3 differential pair of electrodes each on the lower and upper leg, 
referred to electrodes \#14 to \#19 in Figure~\ref{fig:feet}~(a,b,c,d). 
The ankle was subjected to linear, lateral, and angular motions,
where 
flexion, and extension actions were categorized to linear motion, abduction, and adduction were grouped to lateral motion, and inversion, eversion were assigned to angular motion. Hence sEMG signals for three different categories
of actions were acquired and were repeated for 60 experimental trials, from electrode pairs positioned at~\#14~to~\#16. 
Similarly, the knee was exerted to flexion and extension motions, 
and sEMG signals were acquired from positioned electrodes~\#17~to~\#19, for 60 repeated trials.

\begin{figure}[htp!]
\begin{center}
\includegraphics[scale=0.70]{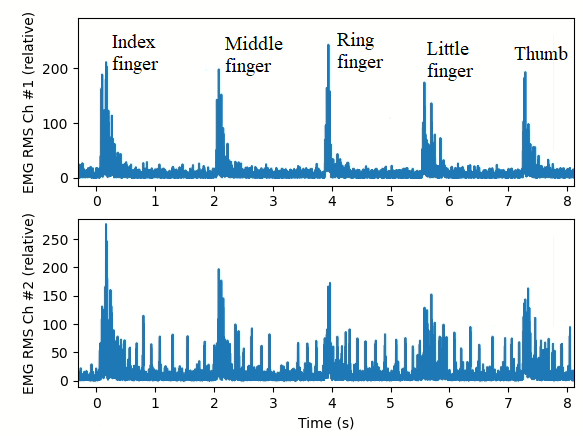}
\caption{RMS response of the acquired sEMG signal for individual finger flexion actions, using two channel sEMG acquisition system.}
\label{fig:emgtwoch}
\end{center}
\end{figure}

EMG signals were recorded for all the actions of upper and lower limb as mentioned above using the multiple channel designed signal acquisition system for training classifier model and validating the same.
Figure~\ref{fig:emgtwoch} shows one such sample of two channel root mean squared (RMS) processed sEMG signal for individual finger flexion actions over time, where RMS was estimated for a 20~ms moving window.
As depicted, the RMS response
of the sEMG signal, showcases a peak value for individual finger actions over time.
The peak amplitude is evidently different for different finger movements for 
signal acquired from both channels.
The channels acquiring sEMG signals that are located at different positions on the forearm, dominantly captures electro-muscular activity generated
by the muscle fibres running beneath the position, else the sEMG signal level is expected to be low.
The varying RMS peak for different finger actions
relates to the dissimilar impact on the same muscle fibres
that are detected by the selected electrode. 
Additionally
electrodes placed at distinct positions shows different RMS peaks
for same finger actions suggesting dissimilar impact of the flexion action on
different group of muscle fibres.
The RMS amplitude shows distinct signature for the origin
of the muscle group signal, and the same was used to classify 
different finger actions.
Similarly distinct pattern of signal was observed on all six channels of the acquired signals for finger and wrist actions, by employing six pair of electrodes around the forearm. 
The action performed by the upper limb was theoretically represented
by a unique vector consisting of EMG signal feature, however the variations involved in
the acquired sEMG signal for same repeated actions were generally observed to be high due to muscle fatigue~\cite{fatigue}, and varying force applied to generate the same actions. Hence the acquired sEMG signals were normalized to 
mitigate the variations mentioned above, instead of using absolute EMG signal.
The moving window RMS value was extracted from the generated sEMG signals for different actions attempted on all the employed channels. The moving window length of 20~ms was calibrated empirically to maximize the energy difference between actuation and relaxation of the muscles. 
The extracted peak RMS value from a channel was considered as a feature, and
the feature vector was generated by collating the same from all the employed channels.
The feature vector defined by $n$ channel was further normalized to reduce vector dimension to $n-1$, by considering the ratio of each RMS peak of $n-1$ channel 
to the one chosen RMS peak of a channel.
The choice of one of the RMS peak values as normalizing constant 
was also a variable, which heavily affects the classifier accuracy, hence
all individual channel RMS peak as normalizing constant
was picked in order to evaluate maximum accuracy for all combinations.
Feature vector of length $n-1$ was then applied to the classifier to predict the upper and lower limb actions, and the same was repeated for $n-1$ times by picking the other channel RMS peak for normalization.
A total of 1980 trials including lower limb and upper limb movements were performed by the same individual at different times,
and the corresponding sEMG signals and its features were extracted and normalized. Support vector machine (SVM) based classifier was considered
the best technique for processing EMG signals as learnt from prior
state of art work~\cite{svm1, svm2, svm3}.
The feature dataset was divided to 80\% training and 
20\% testing classes, and the training features were applied to 
SVM to develop a robust and light weight classifier algorithm, when compared to other neural network models.~\cite{scikit-learn}.

\section{Experimental Results}
\subsection{Finger Flexion Actions}
\begin{figure}[htp!]
\begin{center}
\includegraphics[scale=0.70]{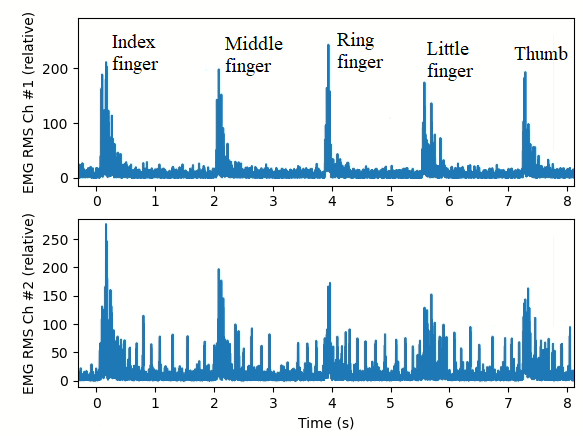}
\caption{EMG signal classification accuracy for 4 fingers with respect to the number of channels used. (*, *, ..) represents the channel combinations corresponding to the validated accuracy.}
\label{fig:emgacc}
\end{center}
\end{figure}

EMG signal was collected at the surface of the skin for all the actions including upper and lower limb mentioned in the previous section.
SVM models for different target classes were attempted sequentially 
starting from upper limb movements of fingers, wrists, elbow, and shoulder joints to lower limb ankle, and knee joint movements.
One such case is the prediction model for 4 finger classes, 
in which the data pertaining to only those 4 finger movements were used to calculate the  accuracy.
Additionally, optimum number and position of electrodes were investigated to achieve maximum classification accuracy for the defined target classes.
The electrodes were positioned and enumerated appropriately, and various electrode combinations extracted accuracy's were compared to identify the 
optimized position of electrodes. 
In the experiment of four finger actions which does not account for thumb movement, features from all combinations of placed electrodes
were employed to estimate accuracy. 
For example, $6_{C1} \times 1$ accuracy was estimated for the
four finger target classes via 1 electrode from the 6 available positioned 
electrodes.
Similarly $6_{C2} \times 2$, and $6_{C3} \times 3$ accuracy were established for the four finger target classes via 2 and 3 electrodes respectively
from the 6 positioned electrodes.
The result of 4 target finger actions with various electrode pairs combinations and associated target classifier accuracy was plotted in Figure~\ref{fig:emgacc}, where each data point (dark blue star) represents a
unique electrode subset combination. The blue line shows the maximum
possible accuracy obtained for the target class for a unique channel
combination, and the same is specified above the line.
The choice of the electrode pairs in the combinations for yielding different target class accuracy depicts a pattern where one or two electrode pairs are common across the target classes. The electrode pairs positioned around~\#3 was the most suitable channel across the target classes, followed by \#4, \#2, \#5, \#1, and \#6 in the order of
preference.
The recommended set of channels in the order suggests the significance of placement of the electrodes pairs. If one has only 2 channels, then it is advised to employ electrode pairs at \#3, and \#4 locations to obtain
maximum accuracy of 75\%.
Similarly if one has 3 channels, then three electrode pairs at \#3, \#4, and \#2 predefined locations is suggested to obtain maximum accuracy.
The predefined positions of electrodes as shown in Figure~\ref{fig:arms}~(a,b) also suggests that most of flexor muscles generally 
run through these electrode positioned at \#3, \#4, and \#2, and hence the 
experimental results for finger flexion are in complete agreement.
A higher signal quality leading to a better classification accuracy was achieved by identifying the target muscle group and placing the electrode
above the identified muscle group.
It was also evident that to classify four target classes involving individual four fingers flexion actions, the SVM model required optimum of four electrodes at
\#2, \#3, \#4, and \#5 positions. More channels does not offer any improvement in accuracy.


\subsection{Optimum Number of Electrodes}
\begin{figure}[htp!]
\begin{center}
\includegraphics[scale=0.60]{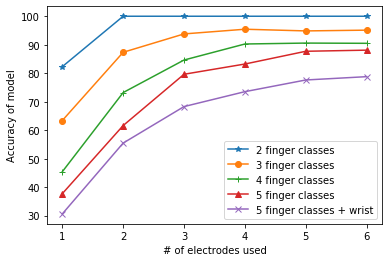}
\caption{sEMG signal based classification accuracy for five finger and wrist targeted individual flexion action versus the number of electrode pairs employed.}
\label{fig:emg_facc}
\end{center}
\end{figure}

\begin{figure}[htp!]
\begin{center}
\includegraphics[scale=0.68]{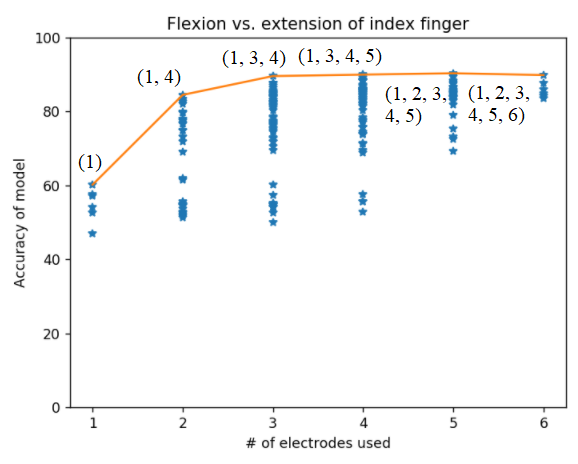}
\caption{sEMG signal classification accuracy for flexion and extension action of index finger with respect to number of electrode pairs.}
\label{fig:emg_flex_ext}
\end{center}
\end{figure}

The maximum classifier accuracy for different number of electrode pairs
was plotted for different target classes
which primarily included flexion actions starting from two individual finger movements to five individual finger motions including thumb, with
an addition of wrist flexion. 
Figure~\ref{fig:emg_facc} shows only the maximum classifier accuracy for a particular electrode pairs arrangement from the 
overall subset of channel combinations.
It is clearly evident that for achieving maximum classifier accuracy, an optimum number of electrode pairs at specific position was adequate. The optimum number of the electrode pairs relates to the distinct number of classes that are targeted to classify.
For differentiating two individual finger actions, two electrode pairs were optimum, and similarly for five finger actions, five electrodes were recommended. 
sEMG signals acquired from six electrode pairs were required to classify between individual fingers and wrists flexion action as shown in the Figure~\ref{fig:emg_facc}.
It was also evident that the maximum accuracy dropped from 100\% to 85\%, when the number of target classes was extended from independent two finger movements to 5 finger actions.
The 6 target classes including five finger flexion actions and wrist flexion movement showcased further drop in classifier accuracy to 75\% by employing all 6 channels of sEMG acquisition system.

sEMG signal classification accuracy for two actions of index finger: flexion,
and extension is shown in Figure~\ref{fig:emg_flex_ext}. A minimum of three electrodes were required to classify between index finger action.
A similar accuracy profile was seen for other fingers actions too.
The highest accuracy represented by data point (blue star) for index finger actions was obtained by employing electrode pairs positioned at \#1, \#3, and \#4.
The orange line marks the connected line for maximum accuracy achieved for different combinations of electrodes when six channel sEMG acquisition system
was employed. 
The index finger action also reinstates the earlier observation on utilizing $n$ channels for classifying between $n$ target classes, with accuracy of 2 channels is very close to the maximum accuracy characterized for optimal 3 channels.

\subsection{Supination and Pronation of hand}
\begin{figure}[htp!]
\begin{center}
\includegraphics[scale=0.75]{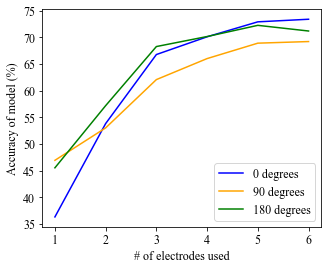}
\caption{ sEMG signal classification accuracy for 5 finger flexion actions
at different forearm postures including supination (0\textdegree) and pronation (180\textdegree) states with respect to the number of electrode pairs.}
\label{fig:emg_rot_acc}
\end{center}
\end{figure}

Supination and Pronation forms the rotational positions of the hand with forearm, and distinguishing hand flexion actions while in different rotational posture of forearm was investigated.
Most of the muscles responsible for the various actions of the hand are cited within the forearm. Since all the 6 pairs of electrodes were placed on the forearm, supination and pronation of the forearm will result in mechanical disruption for the prefixed electrodes and its associated wires. Hence to mitigate this artifact,
sEMG data for all five finger flexion action was collected at three different
angles of the forearm involving 0\textdegree, 90\textdegree, and 180\textdegree.
0\textdegree corresponds to the subject facing the anterior side of the hand, also referred to as supination state, and 180\textdegree correspond to the subject facing the posterior side of the hand and forming
pronation state, and 90\textdegree forms the neutral position of the hand. Figure~\ref{fig:emg_rot_acc} shows the independent classifier accuracy for each of the different posture of forearm.
The data used for testing and training independent classifier model were taken from the same forearm posture's.
The accuracy plot suggests that similar classification accuracy of 73\%, were obtained for finger flexion action when in supination or pronation state. 
The classification accuracy dropped by merely 5\% for in between configuration.

\begin{figure}[htp!]
\begin{center}
\includegraphics[scale=0.75]{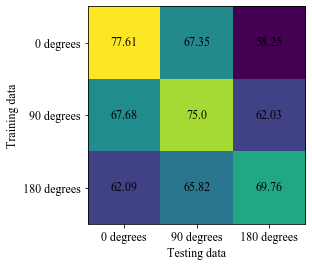}
\caption{Inter-angular posture classification accuracy's for
five finger flexion actions including supination (0\textdegree), neutral (90\textdegree), and pronation (180\textdegree) of the forearm.}
\label{fig:emg_rot_inter}
\end{center}
\end{figure}

Figure~\ref{fig:emg_rot_inter} shows the inter-angular classification accuracy when 6 channels of sEMG acquisition system were used. 
Understandably, data of different angular
posture of the arm were used as testing data to verify finger actions for the current angular posture, leading to a robust characterization results.
It is clearly evident 
that all five finger flexion feature data including testing and training from the same angular posture, presented maximum accuracy as expected. 
The accuracy for finger flexion action drops as the angular difference 
between the postures of trained model and testing model increases.
The experimental result suggests that a system to detect angle of rotation
followed by application of the nearest available model to predict finger action
with acceptable accuracy is possible.
The developed sEMG acquisition system has the capability to characterize 
elbow joint actions to identify the forearm rotation.



\subsection{Actions of Elbow Joint}

\begin{figure}[htp!]
\begin{center}
\includegraphics[scale=1.05]{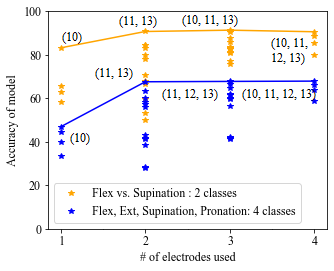}
\caption{sEMG signal based classification accuracy for 
linear and rotational actions of the elbow joint.}
\label{fig:forearm_emg_2ch}
\end{center}
\end{figure}

Various muscles present in the upper arm are responsible for the extension and flexion of the elbow, as well as the pronation and supination of the forearm. Hence a study of these actions and associated sEMG signals were collected to identify and predict these motions. 
Four electrode pairs from the sEMG acquisition system were placed at positions \#10, \#11, \#12, and \#13 across the upper arm as
shown in Figure~\ref{fig:arms}~(c,d), to study the various elbow movements. sEMG signals were acquired for flexion, extension, pronation, and supination of the forearm, by keeping the upper arm fixed.
Two classifier model were generated with one model designed for classifying between linear and rotational motion of elbow joint, and the other model was designed
for classifying between all four actions of elbow joint.
The linear motion includes flexion, and extension actions, and rotational motion represents supination, and pronation movements. 
Figure~\ref{fig:forearm_emg_2ch} shows the classifier accuracy for both models, where {\it Flex}, and {\it Ext} legend represents flexion and extension actions respectively.
High classification results were obtained for identifying actions between rotational, and 
linear actions of the elbow joint. An accuracy of 90\% was characterized using 
two channels for recognizing movements between linear and rotational types.
Between the four electrode pairs positioned at~\#10, \#11, \#12, and \#13, multiple  combinations of electrode pairs including $4_{C1} \times 1$,
$4_{C2} \times 2$, $4_{C3} \times 3$, and $4_{C4}\times4$ were used to identify  optimal number, and position of electrodes to obtain highest classification accuracy of around 90\%.
The electrode pairs placed at \#11, and \#13 were found to offer maximum accuracy and were observed as common positions across 
different electrode groups employed
for classifying between linear and rotational actions, starting from two to four electrode pairs. It is also evident that only two electrode pairs were optimal to 
find the maximum classification accuracy.
Incidentally, the degree of motions generated by elbow in this experiment, matches with the optimal number of electrodes offering maximum accuracy.
Figure~\ref{fig:forearm_emg_2ch} also showcases the classification accuracy 
for identifying actions between four movements that includes flexion,
extension, supination and pronation motion of the elbow joint. 
It was observed that with just two electrode pairs maximum accuracy of 68\% was achieved to classify among four actions. Further improvement in the accuracy of the classifier is possible by fusing both four target and two target 
classifier model.


\subsection{Actions of Shoulder Joint}

\begin{figure}[htp!]
\begin{center}
\includegraphics[scale=0.63]{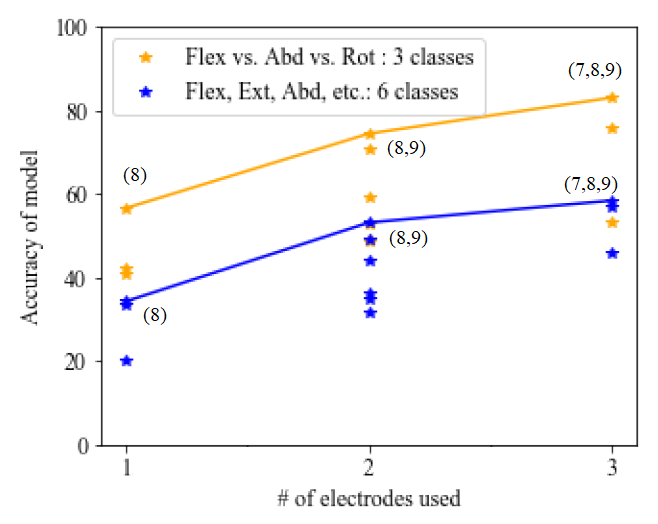}
\caption{sEMG signal based classification accuracy for 3 target classes that includes linear, lateral, and rotational motion of shoulder joint, and 6 target classes
that includes flexion, extension, abduction, adduction, pronation and supination of shoulder joint with respect to the number of electrode pairs used.}
\label{fig:shoulder_3ch}
\end{center}
\end{figure}

The shoulder has several muscles responsible for offering three degrees of motion. Due to the presence of several functional muscles, it was difficult to place as many electrode pairs as the muscle pairs. 
Hence 3 electrode pairs were placed at \#7, \#8, and \#9 positions
across the upper arm as shown in Figure~\ref{fig:arms}~(c,d) to characterize  various shoulder movements. 
The muscles at these positions were observed to wiggle on performing  all actions of shoulder joint, hence these positions were considered optimum for the study on shoulder actions.
sEMG signals were collected for six shoulder actions namely flexion, extension, abduction, adduction, supination, and pronation.
Flexion and extension actions represents linear motion, 
abduction and adduction forms lateral motion, and 
supination and pronation falls into rotational motion.
Figure~\ref{fig:shoulder_3ch} shows the accuracy of classification 
between independent linear, lateral, and rotational actions,
where Flex, Abd, Rot, Ext legend represents flexion, abduction, rotational, and extension actions respectively.
It was observed that with three channels, the system predicted three distinct shoulder actions with ~80\% accuracy. Figure~\ref{fig:shoulder_3ch} also plots the classification accuracy for 6 distinct actions. The maximum accuracy reported was around 60\% for classifying between six shoulder actions.


\subsection{Lower limb}

\begin{figure}[htp!]
\begin{center}
\includegraphics[scale=0.85]{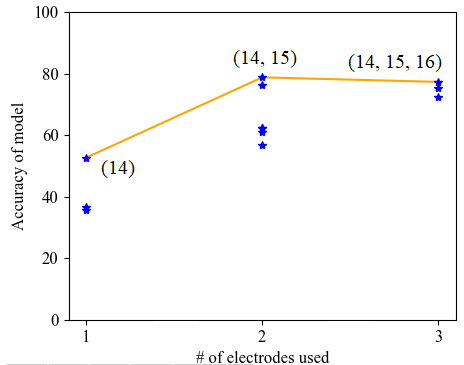}
\caption{sEMG signal driven classifier model showing accuracy for lower limb ankle joint actions between linear, lateral, and angular motions.}
\label{fig:ankle_3cl}
\end{center}
\end{figure}

In the lower limb section, the therapy is generally focused towards the lower extremeties including knee and ankle joints in the early sessions of the rehabilitation, especially for patients suffering from post stroke, fracture surgery, and paralysis~\cite{lower-rehab}.
The ankle movement as a first step is advised for restoring walking ability, and is also useful to prevent patients from suffering with Deep vein thrombosis (DVT)~\cite{vinayDVTjournal, vinayDVTconf}.
Three electrode pairs were placed on the lower leg region along \#14, \#15, and \#16 positions 
of the lower limb as shown in the Figure~\ref{fig:feet}(c,d) to capture calf muscles movement during ankle actuation. With three electrode pairs, all possible combinations were investigated to find the optimized number of channels and best position of electrode pairs to yield maximum classification accuracy between the three class of motions.
The flexion and extension movements are represented as linear motion,
abduction and adduction actions are categorized under lateral 
motion, and inversion and eversion actions fall into angular motion.
A classifier model was generated for predicting three motions, and the accuracy results are reported in Figure~\ref{fig:ankle_3cl}.
It shows that 2 and 3 electrode pairs offered maximum accuracy of around 80\% to distinguish between the three degree of motions for ankle joints. The electrode pairs positioned at \#14 and \#15 were adequate to capture the ankle movements.

\begin{figure}[htp!]
\begin{center}
\includegraphics[scale=0.85]{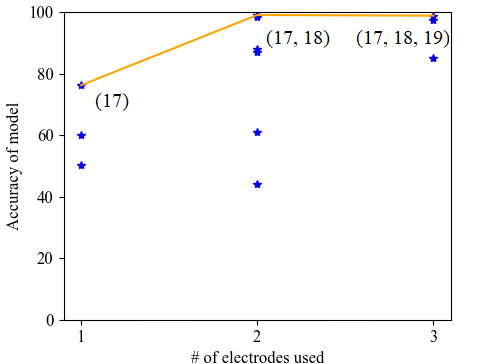}
\caption{sEMG signal driven classifier showing accuracy for knee joint flexion and extension actions.}
\label{fig:knee_3cl}
\end{center}
\end{figure}

Similarly three electrode pairs were placed on the upper leg region along the \#17, \#18, and \#19 positions as shown in the Figure~\ref{fig:feet}~(a,b), and sEMG signals were acquired for
multiple trials, keeping upper leg region fixed, and moving only the lower leg region. 
All possible combinations with single electrode pairs, and multiple electrode pairs were utilized to characterize between flexion and extension actions of the knee joint.  
The knee joint actuation showed maximum accuracy of 100\% to classify between two linear actions with 2 electrode pairings as shown in the Figure~\ref{fig:knee_3cl}. The electrode pairs placed at \#17 and \#18, were found as the optimal positions to acquire sEMG signals and characterize the knee joint actions.
Overall multiple actions of upper and lower limb were identified using different classifier models at acceptable accuracy. 

\section{CONCLUSION}

An in-house 
eight channel sEMG signal acquisition system was designed, fabricated, and was used to characterize major parts of upper and lower limb actions. The developed system with 8 channel differential electrodes were employed to find the best positions to capture 
strong electro-muscular activity signal and yield maximum classification accuracy for different limb movements. 
The optimal position of electrode was recommended to be close to 
the muscle group of interest, to successfully characterize the limb movements.
The optimum number of electrode pairs for classifying different target classes were also investigated with all possible combinations of electrodes and 
classification results achieved from these combinations.
Only a subset of electrode pairs were required to achieve maximum classification accuracy beyond which the accuracy did not improve further.
SVM classifier was adopted for characterizing the movements of upper and lower limb parts including finger, wrists, elbow, shoulder, knee, and ankle joints.
Majority of the upper limb and lower limb motor movements were classified using surface EMG signals adequately using a primitive ML model. 
\bibliographystyle{unsrt}
\bibliography{paper}

\begin{thebibliography}{10}

\bibitem{home}
Sunghoon~I. Lee, Catherine~P. Adans-Dester, Matteo Grimaldi, Ariel~V. Dowling, Peter~C. Horak, Randie~M. Black-Schaffer, Paolo Bonato, and Joseph~T. Gwin.
\newblock Enabling stroke rehabilitation in home and community settings: A wearable sensor-based approach for upper-limb motor training.
\newblock {\em IEEE Journal of Translational Engineering in Health and Medicine}, 6:1--11, 2018.

\bibitem{upperlimb1}
Gert Kwakkel, Boudewijn~J. Kollen, Jeroen van~der Grond, and Arie~J.H. Prevo.
\newblock Probability of regaining dexterity in the flaccid upper limb.
\newblock {\em Stroke}, 34(9):2181--2186, 2003.

\bibitem{upperlimb2}
Leire Santisteban, Maxime Térémetz, Jean-Pierre Bleton, Jean-Claude Baron, Marc~A. Maier, and Påvel~G. Lindberg.
\newblock Upper limb outcome measures used in stroke rehabilitation studies: A systematic literature review.
\newblock {\em PLOS ONE}, 11(5):1--16, 05 2016.

\bibitem{rehab1}
BOUWIEN C~M SMITS-ENGELSMAN, RAINER BLANK, ANNE-CLAIRE VAN DER~KAAY, RIANNE MOSTERD-VAN DER~MEIJS, ELLEN VLUGT-VAN DEN~BRAND, HELENE~J POLATAJKO, and PETER~H WILSON.
\newblock Efficacy of interventions to improve motor performance in children with developmental coordination disorder: a combined systematic review and meta-analysis.
\newblock {\em Developmental Medicine \& Child Neurology}, 55(3):229--237, 2013.

\bibitem{rehab2}
Michelle Wang and Denise Reid.
\newblock Virtual reality in pediatric neurorehabilitation: Attention deficit hyperactivity disorder, autism and cerebral palsy.
\newblock {\em Neuroepidemiology}, 36:2--18, 11 2010.

\bibitem{rehab3}
Ilaria Bortone, Daniele Leonardis, Nicola Mastronicola, Alessandra Crecchi, Luca Bonfiglio, Caterina Procopio, Massimiliano Solazzi, and Antonio Frisoli.
\newblock Wearable haptics and immersive virtual reality rehabilitation training in children with neuromotor impairments.
\newblock {\em IEEE Transactions on Neural Systems and Rehabilitation Engineering}, 26(7):1469--1478, 2018.

\bibitem{management}
Pamela~W. Duncan, Richard Zorowitz, Barbara Bates, John~Y. Choi, Jonathan~J. Glasberg, Glenn~D. Graham, Richard~C. Katz, Kerri Lamberty, and Dean Reker.
\newblock Management of adult stroke rehabilitation care.
\newblock {\em Stroke}, 36(9):e100--e143, 2005.

\bibitem{management1}
Wilma~M. Hopman and Jane Verner.
\newblock Quality of life during and after inpatient stroke rehabilitation.
\newblock {\em Stroke}, 34(3):801--805, 2003.

\bibitem{management2}
Edwin~Daniel Oña~Simbaña, Patricia Sánchez-Herrera~Baeza, Alberto Jardón~Huete, and Carlos Balaguer.
\newblock Review of automated systems for upper limbs functional assessment in neurorehabilitation.
\newblock {\em IEEE Access}, 7:32352--32367, 2019.

\bibitem{running}
Henriette van Praag, Brian~R. Christie, Terrence~J. Sejnowski, and Fred~H. Gage.
\newblock Running enhances neurogenesis, learning, and long-term potentiation in mice.
\newblock {\em Proceedings of the National Academy of Sciences}, 96(23):13427--13431, 1999.

\bibitem{running2}
Hyo~Youl Moon, Andreas Becke, David Berron, Benjamin Becker, Nirnath Sah, Galit Benoni, Emma Janke, Susan Lubejko, Nigel Greig, Julie~A Mattison, Emrah Duzel, and Henriette van Praag.
\newblock Running-induced systemic cathepsin b secretion is associated with memory function.
\newblock {\em Cell Metabolism}, 24, 06 2016.

\bibitem{exercise}
Hiroshi Maejima, Shuta Ninuma, Akane Okuda, Takahiro Inoue, and Masataka Hayashi.
\newblock Exercise and low-level gabaa receptor inhibition modulate locomotor activity and the expression of bdnf accompanied by changes in epigenetic regulation in the hippocampus.
\newblock {\em Neuroscience Letters}, 685:18--23, 2018.

\bibitem{exercise2}
Beth~E. Fisher, Giselle~M. Petzinger, Kerry Nixon, Elizabeth Hogg, Samuel Bremmer, Charles~K. Meshul, and Michael~W. Jakowec.
\newblock Exercise-induced behavioral recovery and neuroplasticity in the 1-methyl-4-phenyl-1,2,3,6-tetrahydropyridine-lesioned mouse basal ganglia.
\newblock {\em Journal of Neuroscience Research}, 77(3):378--390, 2004.

\bibitem{quantitative}
Chen Wang, Liang Peng, Zeng-Guang Hou, Jingyue Li, Tong Zhang, and Jun Zhao.
\newblock Quantitative assessment of upper-limb motor function for post-stroke rehabilitation based on motor synergy analysis and multi-modality fusion.
\newblock {\em IEEE Transactions on Neural Systems and Rehabilitation Engineering}, 28(4):943--952, 2020.

\bibitem{frequency-range}
Nurhazimah Nazmi, Mohd~Azizi Abdul~Rahman, Shin-ichiroh Yamamoto, Siti Ahmad, Hairi Zamzuri, and Saiful Mazlan.
\newblock A review of classification techniques of emg signals during isotonic and isometric contractions.
\newblock {\em Sensors}, 16:1304, 08 2016.

\bibitem{frequency-range2}
Laxmi Shaw and Sangeeta Bhaga.
\newblock Online emg signal analysis for diagnosis of neuromuscular diseases by using pca and pnn.
\newblock {\em International Journal Of Engineering Science and Technology 0975-5462}, 4:4453--4459, 10 2012.

\bibitem{surface}
T.~{Koshio}, S.~{Sakurazawa}, M.~{Toda}, J.~{Akita}, K.~{Kondo}, and Y.~{Nakamura}.
\newblock Identification of surface and deep layer muscles activity by surface emg.
\newblock In {\em 2012 Proceedings of SICE Annual Conference (SICE)}, pages 1816--1821, Aug 2012.

\bibitem{realtime}
V.~{Glaser}, A.~{Holobar}, and D.~{Zazula}.
\newblock Real-time motor unit identification from high-density surface emg.
\newblock {\em IEEE Transactions on Neural Systems and Rehabilitation Engineering}, 21(6):949--958, Nov 2013.

\bibitem{artifact}
O.~W. {Samuel}, X.~{Li}, P.~{Fang}, and G.~{Li}.
\newblock Examining the effect of subjects' mobility on upper-limb motion identification based on emg-pattern recognition.
\newblock In {\em 2016 Asia-Pacific Conference on Intelligent Robot Systems (ACIRS)}, pages 137--141, July 2016.

\bibitem{motor-neuron}
M.~{Chen}, X.~{Zhang}, and P.~{Zhou}.
\newblock A novel validation approach for high-density surface emg decomposition in motor neuron disease.
\newblock {\em IEEE Transactions on Neural Systems and Rehabilitation Engineering}, 26(6):1161--1168, June 2018.

\bibitem{poststroke}
X.~{Li}, A.~{Holobar}, M.~{Gazzoni}, R.~{Merletti}, W.~Z. {Rymer}, and P.~{Zhou}.
\newblock Examination of poststroke alteration in motor unit firing behavior using high-density surface emg decomposition.
\newblock {\em IEEE Transactions on Biomedical Engineering}, 62(5):1242--1252, May 2015.

\bibitem{myoband}
{\em MyoBand}, 2018 (accessed Jan 15, 2020).

\bibitem{natus}
{\em Nicolet EDX EMG / NCS / EP / IOM System}, 2020 (accessed Dec 15, 2020).

\bibitem{roam}
{\em ROAM Wireless EMG system}, 2020 (accessed Dec 30, 2020).

\bibitem{neurosoft}
{\em Neuro-MEP-8}, 2020 (accessed Dec 12, 2020).

\bibitem{delsys}
{\em Trigno Sensors}, 2020 (accessed Dec 11, 2020).

\bibitem{cometa}
{\em Mini Wave Infinity}, 2020 (accessed Dec 13, 2020).

\bibitem{sharmilaICPR}
Sharmila Mani and Madhav Rao.
\newblock Feasibility study of using myoband for learning electronic keyboard.
\newblock In {\em 2020 25th International Conference on Pattern Recognition (ICPR)}, pages 10196--10202, 2021.

\bibitem{myoband1}
Nikitha Anil and S.H Sreeletha.
\newblock Emg based gesture recognition using machine learning.
\newblock In {\em 2018 Second International Conference on Intelligent Computing and Control Systems (ICICCS)}, pages 1560--1564, 2018.

\bibitem{vinay-sEMG}
V.~{Chandrasekhar}, V.~{Vazhayil}, and M.~{Rao}.
\newblock Design of a real time portable low-cost multi-channel surface electromyography system to aid neuromuscular disorder and post stroke rehabilitation patients.
\newblock In {\em 2020 42nd Annual International Conference of the IEEE Engineering in Medicine Biology Society (EMBC)}, pages 4138--4142, 2020.

\bibitem{emgpattern}
A.~{Zhang}, N.~{Gao}, L.~{Wang}, and Q.~{Li}.
\newblock Combined influence of classifiers, window lengths and number of channels on emg pattern recognition for upper limb movement classification.
\newblock In {\em 2018 11th International Congress on Image and Signal Processing, BioMedical Engineering and Informatics (CISP-BMEI)}, pages 1--5, Oct 2018.

\bibitem{sapsanis2013}
C.~Sapsanis.
\newblock {\em Recognition of basic hand movements using Electromyography}.
\newblock PhD thesis, Dept. Elect. and Comp. Engg, Universtity of Patras, 06 2013.

\bibitem{sapsanis201306}
C.~Sapsanis, G.~Georgoulas, and A.~Tzes.
\newblock Emg based classification of basic hand movements based on time-frequency features.
\newblock In {\em 21st Mediterranean Conference on Control and Automation}, pages 716--722, 2013.

\bibitem{sapsanis201307}
C.~Sapsanis, G.~Georgoulas, A.~Tzes, and D.~Lymberopoulos.
\newblock Improving {EMG} based classification of basic hand movements using {EMD}.
\newblock In {\em IEEE Engineering in Medicine and Biology Society Conference}, volume 2013, pages 5754--5757, 07 2013.

\bibitem{saponas2008}
T.~Saponas, D.~Tan, D.~Morris, and R.~Balakrishnan.
\newblock Demonstrating the feasibility of using forearm electromyography for muscle-computer interfaces.
\newblock In {\em Conference on Human Factors in Computing Systems - Proceedings}, pages 515--524, 01 2008.

\bibitem{Saponas2009}
T.~Saponas, D.~Tan, D.~Morris, R.~Balakrishnan, J.~Turner, and J.~Landay.
\newblock Enabling always-available input with muscle-computer interfaces.
\newblock In {\em UIST 2009 - Proceedings of the 22nd Annual ACM Symposium on User Interface Software and Technology}, pages 167--176, 01 2009.

\bibitem{feature}
M.~S. {Hazam Majid}, W.~{Khairunizam}, A.~B. {Shahriman}, I.~{Zunaidi}, B.~N. {Sahyudi}, and M.~R. {Zuradzman}.
\newblock Emg feature extractions for upper-limb functional movement during rehabilitation.
\newblock In {\em 2018 International Conference on Intelligent Informatics and Biomedical Sciences (ICIIBMS)}, volume~3, pages 314--320, Oct 2018.

\bibitem{rms-window}
Haopeng Wang, Kiriaki~J. Rajotte, He~Wang, Chenyun Dai, Ziling Zhu, Moinuddin Bhuiyan, Xinming Huang, and Edward~A. Clancy.
\newblock Optimal estimation of emg standard deviation (emg $\sigma$ ) in additive measurement noise: Model-based derivations and their implications.
\newblock {\em IEEE Transactions on Neural Systems and Rehabilitation Engineering}, 27(12):2328--2335, 2019.

\bibitem{rms1}
Thiago Fukuda, Jorge Echeimberg, José Pompeu, Paulo Lucareli, Silvio Garbelotti~Junior, R.O. Gimenes, and A.~Apolinário.
\newblock Root mean square value of the electromyographic signal in the isometric torque of the quadriceps, hamstrings and brachial biceps muscles in female.
\newblock {\em Applied Research}, 10:32--39, 01 2010.

\bibitem{rms2}
Madeleine Lowery and Mark O'Malley.
\newblock Analysis and simulation of changes in emg amplitude during high-level fatiguing contractions.
\newblock {\em IEEE transactions on bio-medical engineering}, 50:1052--62, 10 2003.

\bibitem{github}
surface-emg data.
\newblock \url{https://github.com/7andahalf/sEMG}.

\bibitem{armfig}
Andrew~F. Currier.
\newblock In {\em (Ed.) The Foundation Library (New York, NY: The Educational Society)}, 1911.

\bibitem{fatigue}
M.~D.~F. {Ma'as}, {Masitoh}, A.~Z.~U. {Azmi}, and {Suprijanto}.
\newblock Real-time muscle fatigue monitoring based on median frequency of electromyography signal.
\newblock In {\em 2017 5th International Conference on Instrumentation, Control, and Automation (ICA)}, pages 135--139, Aug 2017.

\bibitem{svm1}
Yogesh Paul, Vibha Goyal, and Ram~Avtar Jaswal.
\newblock Comparative analysis between svm knn classifier for emg signal classification on elementary time domain features.
\newblock In {\em 2017 4th International Conference on Signal Processing, Computing and Control (ISPCC)}, pages 169--175, 2017.

\bibitem{svm2}
Hazirah Hasni, Norashikin Yahya, Vijanth~S. Asirvadam, and Munsif~Ali Jatoi.
\newblock Analysis of electromyogram (emg) for detection of neuromuscular disorders.
\newblock In {\em 2018 International Conference on Intelligent and Advanced System (ICIAS)}, pages 1--6, 2018.

\bibitem{svm3}
Diana Toledo~Pérez, Juvenal Rodriguez, Roberto Gómez~Loenzo, and Juan Jauregui.
\newblock Support vector machine-based emg signal classification techniques: A review.
\newblock {\em Applied Sciences}, 9:4402, 10 2019.

\bibitem{scikit-learn}
F.~Pedregosa, G.~Varoquaux, A.~Gramfort, V.~Michel, B.~Thirion, O.~Grisel, M.~Blondel, P.~Prettenhofer, R.~Weiss, V.~Dubourg, J.~Vanderplas, A.~Passos, D.~Cournapeau, M.~Brucher, M.~Perrot, and E.~Duchesnay.
\newblock Scikit-learn: Machine learning in {P}ython.
\newblock {\em Journal of Machine Learning Research}, 12:2825--2830, 2011.

\bibitem{lower-rehab}
R.~{Xu}.
\newblock A novel design of the lower limb rehabilitation robot.
\newblock In {\em 2020 6th International Conference on Mechanical Engineering and Automation Science (ICMEAS)}, pages 254--259, 2020.

\bibitem{vinayDVTjournal}
K.~{Vinay}, K.~{Nagaraj}, H.~R. {Arvinda}, V.~{Vikas}, and M.~{Rao}.
\newblock Design of a device for lower limb prophylaxis and exercise.
\newblock {\em IEEE Journal of Translational Engineering in Health and Medicine}, 9:1--7, 2021.

\bibitem{vinayDVTconf}
C.~K. {Vinay}, V.~{Vazhiyal}, and M.~{Rao}.
\newblock Design of a non-invasive pulse rate controlled deep vein thrombosis prophylaxis lower limb device.
\newblock In {\em 2019 41st Annual International Conference of the IEEE Engineering in Medicine and Biology Society (EMBC)}, pages 5407--5410, 2019.

\end{thebibliography}

\end{document}